# Attack-Defense Quantification Based On Game-Theory

Su Yang, Yuqing Zhang, Chensi Wu

*National computer Network Intrusion Protection Center(University of Chinese of Academy of Sciences), Beijing 101408*

**Abstract –** With the developing of the attack and defense technology, the cyber environment has been more and more sophisticated. We failed to give an accurate evaluation of network security situation, as we lack a more accurate quantitative evaluation of attack-defense behaviors. In response to this situation, we proposed an attack-defense stochastic game model (ADSGM), analyzed the different security property of distinct defense mechanism, and put forward a corresponding utility calculation coping with the distinct defense mechanism. Through a case study, we showed the impact of active defense and the risk of attack exposure, demonstrated the effectiveness of our methods on attack-defense behavior quantification. This paper filled with the gap in the quantitative assessment of defensive measures, to make the quantitative evaluation of attack-defense more comprehensive and accurate.

# Introduction

With the developing of the attack and defense technology, the cyber environment has been so sophisticated that the network administrators or experts cannot accurately evaluate the security situation of a network system on intuition. On the other hand, a wide variety of threats continually cause a huge damage to the governments and enterprises both in economy and reputation. A major reason for this dilemma is that the defenders lack a more accurate quantitative evaluation of attack-defense behaviors. In addition, the attack tricks are more and more sophisticated and automatic.

In a cyber environment, the attack-defense behaviors are a series of actions the adversaries and the defenders take for different intentions. As the adversaries and the defenders confront each other continually, the interaction between the attack behaviors and the defense behaviors is a game process. Different behavior choices mean the different payoffs which include benefits and costs, and this produces the attack-defense strategy that profiles the preference. By studying the interaction between attack and defense behaviors, the defenders are able to produce a reasonable optimal defense strategy. What's more, when the attack-defense game is in equilibrium, the corresponding payoff shows the maximum security risk value of the target network, and this value equals to the network security situation. Particularly, at the moment that the adversary choose the optimal attack strategy and the costs still exceed the benefit, the network can be regarded as safe.

There are a lot of metrics and evaluation approaches that try to evaluate the security situation of the network both in qualitative and quantitative. Attack graph[1] is a typical work, which applies topological graph to analyze the security vulnerabilities and the interaction of nodes. It can be used to evaluate the cybersecurity of enterprise architectures[2], [3]. However, it only considers the vulnerabilities and the relationship between nodes, ignoring the other threats like spoofing attack and hijack attack, furthermore, it neglects the impact of defensive measures. A Bayesian network[4]–[6] is another popular method, which is used to calculate the uncertainty and inherent uncertainty within cyber environment. For example, [5] applied Bayesian networks on attack success rate whereas [4] used it to describe safety risk factors in target network.

Game theoretic evaluation approach [7]–[9] is a comprehensive method which considers both the adversary and the defender who influenced the security situation. As a theory on the interaction of strategies made by multiple decision-makers, game theory has been widely used in network security[10]–[15]. It is suitable to handles the interaction between attackers and defenders and helps the defenders make a more reasonable decision.

However, despite there are a lot of related work evaluating network security situation by game theory, existing work in this area suffers from three key limitations, which lead to the inaccuracy of the evaluation. First, the calculation formula of payoff is oversimplified and impractical, therefore it cannot profile the utility of behaviors clearly. For example, previous work[16] only takes account of recovery capability of defense, the deceptive capability and the tracing capability do not include, so the impact of defense measures is inaccurate. Besides, proactive defense technologies such as cyber deception[17] come into vogue in network protection, the previous works, however, only consider the passive defense and the recovery capability. In fact, many defensive measures like cyber deception, intrusion detection provide two extra capability, which can greatly improve the security of the network. Thirdly, the increasing Advanced Persistent Threat events give us more thought of the length of attack time, the concealment and the scalability of attack should be considered in the utility of behaviors.

This paper adopts and extends the game theoretic methods to evaluate the attack-defense behaviors on different scenarios, by generating the optimal attack strategies, then calculates the maximum security risk value of the target network, which considered as the security situation of the network. The major contributions of this study are stated as follows:

## Contributions:

1) Firstly, we distinguish the difference between the proactive defense and the passive defense in the area of utility calculation.
2) Then, we subdivided the capability of defense measures, and we provide two extra capability "Deceptive Capability" and "Tracing Capability" to the evaluation of attack-defense behaviors.
3) Last but not the least, we tentatively put forward a concept "attack time" in the evaluation of attack-defense behaviors and discuss how it works on the utility calculation.

# Model Overview

To evaluate the attack-defense behaviors and relevant security situation of a target network, this paper summarized the following features of the attack-defense process:

1) Attack-Defense Process is a phased process, and in each phase, the adversary and the defender are going to play a game.

2) The attack-defense behaviors lead to the change of network system state.

3) The uncertainty of attack-defense behaviors: Both the adversaries and the defenders would take more than one actions equally in a certain system state, so the adverse actions are invisible or part of invisible.

4) If the adversaries exposure, they would be punished, so they take account of clearing their traces and retreating from the compromised computers within a certain time, even though they did not achieve their goal.

A stochastic game, which belongs to the dynamic game, is a suitable model, it starts with an initial state, and then it would go through a series of state transformation based on both the actions taken by each player and the current state.

# 1. Attack-Defense Stochastic Game Model

Firstly, in the real cyber environment, there are multiple attackers and multiple defenders, however, this paper try to find the maximum security risk value of the target network, and multiple players have no effect on this, so we make a reasonable assumption to simplify the problem:

**Assumption 1**：We consider multiple attackers as one attacker, and multiple defenders as one defender.

**Definition 1**：Attack-Defense Stochastic Game Model (ADSGM)= {N,S,A,D,P,U} is a zero-sum Stochastic game:

1) **Participant set** N={Attacker, Defender} represents the players in the game;
2) **System state space** S= {$S_0$, $S_1$, $S_2$ ⋯ $S_K$}, $\forall k \in Z$ and $k \geq 0$ consists of all possible state of the target network system. In particular, we regard the target network as a system.
3) **Adversary actions set** A= {$a_1$, $a_2$, $a_3$ ⋯ $a_n$}, where $\emptyset \in A$, and at the state k, the adversary actions set is $A^k$;
4) **Defender actions set** D= {$d_1$, $d_2$, $d_3$ ⋯ $d_m$}, where $\emptyset \in D$, and at the state k, the defender actions set is $D^k$;
5) **Transition probability set** P=S×A×D×S->[0,1]; where $p_{ij}^{kl}(a_i, d_j) \geq 0 \in P$ is the probability that under the actions $a_i$ and $d_j$, the state k transfers to state l.
6) **Utility set** $U_k$: S×A×D×S->$v_k$, where $k = a_i, d_j$ and $v_k$ is the utility value of the players in the game; In addition, since ADSGM is a zero-sum, it means that the sum of the adversary utility and the defender utility is zero: $U_a$ + $U_b$ = 0;
7) **Proof of Nash equilibrium existence:** ADSGM can be regarded as an extension of matrix game for each state, the state S, the Adversary actions A and Defender actions D and the utility value $U_a$ and $U_b$ are finite. As a result, ADSGM is a finite stochastic game and since Fink[18] draw a conclusion that every finite stochastic game exits a stable Nash equilibrium, ADSGM must have a stable Nash equilibrium.

# 2. Attack-Defense scenarios

Network intrusion process consisted of many distinguish stages. For each of the stages, the adversary achieved corresponding tasks, which most of the cases were the condition of the next task. The adversary adopted different attack technique in different stages and the defenders took the corresponding defensive measures, so the confrontation and the game between attack techniques and defense techniques distributed in different stages so we classified them into these stages to produce the strategies. We followed the categories of colasoft[19] in this paper.

Moreover, as Table 1 shown, we divided the defensive measures into two categories: proactive defense and passive defense. It was mainly because that they differed in computing methods for the utility.

Table1：Categories of attack and defense technologies

| Attack-defense Stage | Attack techniques | Passive defense | Proactive defense |
|---|---|---|---|
| **Information collecting stage** | Information collection | Gateway access Control Policy、Intrusion detection、firewall rejection | Mimic Defense; fake devices and services; Virtual topology |
| | Scan | | |
| | Network monitor | anomaly detection; encrypted communication | |
| **Intrusion stage** | Spoofing attack：IP spoofing, phishing email, Watering Hole | Two-Factor Authentication, antivirus gateway, antivirus software、Network Reputation Database, Security Threat Intelligence | Attack flow diversion, Topology restructuring |
| | DOS | Anti-DOS devices and services, filtering policy | Network traffic diversion |
| | Hijack Attack | VPN, SSL Gateway, DNS protection | |
| | Vulnerability Exploitation | Intrusion Detection System, Firewall system, vulnerability scanning services, patch management system | Honey-patches |
| | Malware and malicious code | log analysis, Sandbox dynamic detection, honeypot | |
| | Password attack | Limited login, intensive password policy, uniform identity authentication, digital certificate, weak password scan | Uvauth |
| **Privilege elevation stage** | Vulnerability Exploitation | Intrusion Detection System, Firewall | Honey-patches |

| | | system, vulnerability scanning services, patch management system | |
|---|---|---|---|
| **Lateral transfer stage** | Intranet reflection Attack, Domain penetration, host penetration | Intranet access Control, client security detection | |
| | Certificate theft, Bypass the hash | | Honeywords |
| **Persistent resident stage** | backdoor | Antivirus software, gateway flow analysis | |
| **Tracks eraser stage** | erasing invasion, Ransomware | network audit system, network traffic analysis, attack source traceback | |

# 3. Attack-Defense State Transition of a stochastic game

This paper converted the stochastic game state transition (which means the state transition of the network system) as the state transition of each node in the target network. Since a state transition of a node usually means that the other nodes are going to face new security threats such as the watering hole, hijack attack and so on, and consequently the network system security situation may have changed accordingly. So essentially, the system state transition was owing to the state transitions of each node and this equivalent conversion was reasonable.

In ADSGM, the attack-defense state transition can be modeled as a directed graph G=(S, E), in which S represents the game state set and E describe the transition of state.

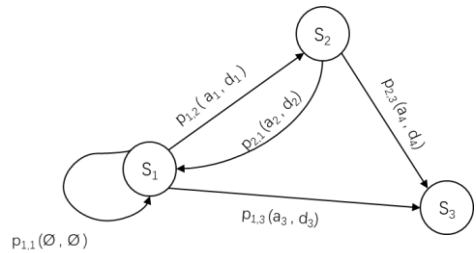

Fig1. Example of ADSGM's state

According to fig.1, the system state space is S = {$S_1$, $S_2$, $S_3$} and the $p_{ij}(a_i, d_j)$ represents the possibility from current state to the next state while the adversary and the defender take the actions $a_i$ and $d_j$ respectively, since not all of the states can transfer certainly to each other, and some transition between the nodes can be impossible.

In addition, for each node, the possible attack-defense actions according to table 1 determined the type of state the node include, and we conclude four types of state for node: a) no privilege; b) remote access privilege; c) root privilege; d) data leak state. Moreover, despite a node could have many types of state, but the adversary tends to let the node into a more dangerous state to obtain maximum payoff, so this paper produced the system state set by a greedy generation algorithm, which chose the most dangerous state of all possible state by a heuristic method. An equally important reason we found was that as the number of states for the node grew, the complexity of ADSGM increased exponentially, so we give the greedy algorithm:

**Algorithm 1. System State Generation Greedy Algorithm**

```
INPUT:  Network topology graph
 G = {N, E},  N_i = {N_near, A^{N_i}, D^{N_i}, S^{N_j}}
OUTPUT: Target system state set: S
```
1. Find all entrance node in N and label as $E_i$ in Entrance[];
2. For $E_i$ in Entrance[ ]:
3.     Initialize $S_0$, N[ ], S[ ]
4.     S[ ]<—$S_0$;
5.     N[ ]<—$E_i$;
6.     For $N_i$ in N[ ]:
7.        each $N_j$ in $N_{near}$[ ]:
8.        Generate all possible actions set $A^{N_j}$ and corresponding utility
9.        Choose the maximum utility action and corresponding state $S^{N_j}$;
10.        S[ ]<—$S^{N_j}$;
11.        N[ ]<—$N_j$;
12.     End for
13. End for

## 4. Utilities of actions in ADSGM

In ADSGM, the attack-defense actions described the attack-defense behaviors, and in the rest of this article, we use both of them indiscriminately.

The attack-defense utility profiled the expectation of an action the adversary and the defender took, and it determined the generation of strategies. Regarding the evolution of attack-defense behaviors as a game of the adversary and the defender, and at the principle that both the adversary and the defender were rational, then we calculated the maximum payoff of all strategies to regard as security situation of the network.

**-Adversary knowledge and utility**

Following the previous researches[7], [16], this paper divided the utility of an attack behavior into two part: one part was the benefit of the attack behavior, it came from the damage the system suffered, and the other one was the costs of the attack behavior, so the attack utility formula was as follow:

**Definition 2:** The attacker utility functions were:

$$U_a = B_a - C_a \qquad (1)$$

Where $B_a$ represented the benefit of the attack behavior and $C_a$ was the cost of the attack behavior.

Furthermore, this paper subdivided the cost part into two type:

a) The cost of the resource, for example, the computing resource and network resource;
b) The risk of exposure refer to the risk of exposing the adversaries themselves.

We found that the second part of costs is the major factor that influences the decision of adversaries, so simplified the problem that the risk of exposure equaled the total cost of the attack action. In the next section, we demonstrated how it affected the strategies and the result of a game.

**- Defensive measures and utility**

In a real network system, the administrator arranges a variety of defensive measurement to hinder the attack behaviors, and consequently prevent the achievement of the adversary goal, and the degree of system security depends on the intention of attack the system can resist, so it is important to measure this ability.

In the previous section, we discussed the attack-defense scenarios and divided the defensive measures into two categories. The basis of categories were not only the different mechanisms they possessed, more importantly, but also we found that they provided different defensive abilities. Traditional passive defensive measures such as firewall, patch management provided the "Recovery Capability", in contrast, proactive defensive measures like cyber deception tried to conceal the target network, create uncertainty and confusion against the adversary's efforts to establish situational awareness and to influence and misdirect adversary perceptions and decision processes. This paper characterized this ability as "Deceptive Capability".

Finally, there was another ability called "Tracing Capability", which often was ignored by evaluation tools. Actually, it was an implicit factor influenced the evolution of attack-defense behaviors, since it related to the cost of attack actions. Higher tracing capability brought a higher risk of exposure, and we try to use a metric called "average attack time" to weight the balance between them.

### -Average attack time and utility

In order to evaluate the risk of exposure for the adversary and the tracing capability of the target system, we tentatively introduced the concept of the attack time and average attack time to calculate the effect of them on utility. Firstly we give the definition of the attack time and the average attack time:

**Definition 3**：*Attack time* was the length of time that the adversary launched a specific attack until they finished.

**Definition 4:** *Average attack time* was the average duration of a specific attack for most adversaries.

The attack time describe the For our evaluation, the average attack time is an empirical constants,

This paper explained how these three capabilities worked on the utility calculation and the evaluation of the network security situation. To begin with, we put forward three attributes to signify the capability of each defensive measures, and introduced the calculation methods:

**Recovery Capability (R)：**

R was corresponding to the loss the system suffered by an attack action, and the recovery capability can mitigate this loss, so we weighted the mitigation by the formula:

$$L' = L - R \quad (2)$$

**Deceptive Capability (E):**

The effect on increasing network security was that it created uncertainty of attack actions, so despite the damage of the attack action was fixed, it could mitigate the damage that the system or device suffered by this action. According to this, we calculated this capability by a payoff matrix:

$$U'_f = P(f|\varphi, \tilde{f})U_f = \frac{\varphi_{f,\tilde{f}}}{N_{\tilde{f}}} U_f \quad (3)$$

Where $f$ presented the true fingerprint whereas $\tilde{f}$ presented the observation fingerprint, $N_{\tilde{f}}$ was the total number of the devices in the system whose observation fingerprint was $\tilde{f}$ and $\varphi_{f,\tilde{f}}$ was the defense configuration describing the number of the true fingerprint($f$)which observed as $\tilde{f}$.

**Tracing Capability (T):**

The tracing capability described the capability of discovering the attack actions, and this paper formalized it as a time-related function:

$$T(t) = \alpha * T/t \quad (4)$$

Where α is empirical coefficient and T is a constant represent "Attack time" of a certain attack action.

Then we extended the utility calculation formula based on (1), (2) and (4).

**Definition 4:** The adversary utility functions were:
$$U_{a,d} = (L - R) * V - T(t)$$
$$= (L - R) * V - \alpha T/t \quad (5)$$
V represented the asset value and L can be divided into C for confidentiality, I for integrity and A for availability.

**-Calculating the payoff matrix**

ADSGM was a zero-sum game, and we regarded it as a matrix game with a Markov decision process[20], there was an example of a matrix game at state $S_k$:

Table2: Example of a payoff matrix

| $S_k$ | $a_1$ | $a_2$ | $a_3$ |
|---|---|---|---|
| $d_1$ | $s^k_{a_1,d_1}$ | $s^k_{a_2,d_1}$ | $s^k_{a_3,d_1}$ |
| $d_2$ | $s^k_{a_1,d_j}$ | $s^k_{a_2,d_2}$ | $s^k_{a_3,d_2}$ |
| $\emptyset$ | $s^k_{a_1,d_j}$ | $s^k_{a_2,d_3}$ | $s^k_{a_3,d_3}$ |

Each column of the table corresponding to the possible attack actions and each row shows the possible defense actions. The element $s^k_{a_i,d_j}$ represented the payoff of the adversary, and the payoff of defenders was $-s^k_{a_i,d_j}$ because of the zero-sum game.

In ADSGM, the payoff matrix elements were as follow:
$$s^k_{ij} = u^k_{ij} + \sum_{l=1}^{k} p^{kl}_{ij}(a_i, ed_j)s_l \ [20] \quad (6)$$
Where $p^{kl}_{ij}(a_i, d_j) \geqslant 0$ was the transition probability from state k to state l. $u^k_{ij}$ was the utility of adversary utility, $\sum_{l=1}^{k} p^{kl}_{ij}(a_i, d_j)s_l$ represented the indirect benefits, and the value of the game was:
$$v_k = Val(S_k) \quad (7)$$

## Solving ADSGM Algorithm

This paper followed the previous study[20] using Shapley iterative algorithm to solve formula (6), and the existence theorem of game value has been given which would not be repeated here. Now giving the algorithm:

**Algorithm 1. Attack-Defense Stochastic Game Algorithm**

INPUT：ADSGM = {N,S,A,D,P,U} and δ
OUTPUT：Optimal attack and defense strategies.
1. Initialize $v^0 = (v^0(1), v^0(2), \cdots, v^0(K))$;
2. Repeat
3.    For each $S_k \epsilon S$ do
4.      For all $s^k_{ij}$ do
5.        Replace $s_l$ in (7) as $v_l$;
6.      end for
7.      Calculate $v^{r+1}_k = Val(u^k_{ij} + \sum_{l=1}^{k} p^{kl}_{ij}(a_i, d_j)v^r_l)$;
8.    end for
9.    for each $S_k \epsilon S$ do
10.      $v_k \leftarrow Val(S_k)$;
11.    end for
12. until $|v^{r+1}_k - v^r_k| < \delta, \forall S_k \epsilon S$
13. for each $S_k \epsilon S$ do
14.    $(\pi^k_a, \pi^k_d) = Solve(S_k)$;
15. end for
16. return $(\pi^k_a, \pi^k_d)$.

# Case study

First, this paper supposed that there was a typical network system topology as fig2. The adversary could access client A and B by the Internet, and there was a firewall between the Internet and the intranet. The access rules of the firewall were as table 2 listed. In particular, the shadow asset was the "copy" of the key asset and it misdirected adversary perceptions and decision processes.

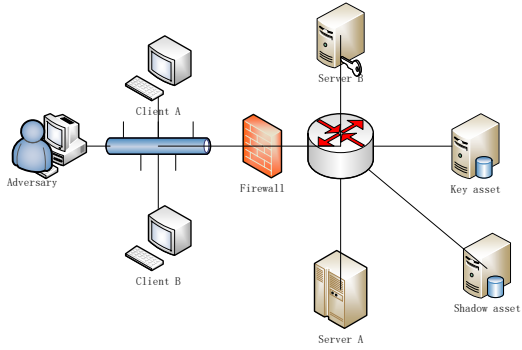

Fig.2 Topology of the example network

Table2: Access Control Rules of Firewall

| From | To | Access Control |
|---|---|---|
| All | Server B | Allow |
| Client B | Server A | Allow |
| Client A | Client B | Allow |
| Server B | key asset | Allow |
| Server B | Shadow asset | Allow |

Then, this paper listed all possible attack and defense actions on the target network based on Table1:

Table3: Adversary Actions List

| A | describe |
|---|---|
| $a_1$ | weak password attack |
| $a_2$ | CVE |
| $a_3$ | DOS |
| $a_4$ | CVE |
| $a_5$ | malicious code |
| $a_6$ | phishing email |
| $a_7$ | Bypass the hash |
| $a_8$ | CVE |

Table4: Defender Actions List

| D | describe |
|---|---|
| $d_1$ | Limited login |
| $d_2$ | anti-virus software |
| $d_3$ | Patch manager system |
| $d_4$ | firewall |
| $d_5$ | Ø |

Besides, this paper supposed the adversary actions sets and defense configuration were as follow:

Table5: Adversary actions and defense configuration of each node

| Node | Action set | Attack effect |
|---|---|---|
| Client A | $a_2$ | root |
| | $a_3$ | no privilege |
| | $a_6$ | user |
| Client B | $a_1$ | user |
| | $a_3$ | no privilege |
| | $a_4$ | user |
| Server A | $a_2, d_3$ | user |
| | $a_5, d_2$ | user |
| Server B | $a_1, d_1$ | root |
| | $a_6, d_4$ | user |
| Key Asset | $a_7$ | user |
| Shadow Asset | $a_7$ | user |

Now, we assumed that the adversary had compromised client A, and recognized client A as an entrance node, then generated the system state set by Algorithm 1 and output the directed graph fig.3.

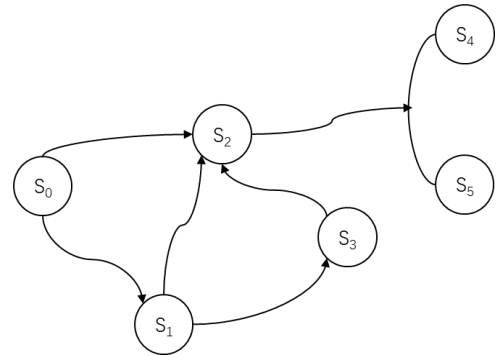

Fig.3 Attack-defense stochastic game state graph

$S_0$ was the system state that client A had been compromised, the adversary actions set was $A^0 = \{a_1, a_2, a_3, a_4\}$ and the defender actions set $D^0 = \{d_1, d_5\}$, where especially, $d_5$ is Ø

| $S_0$ | | $d_1$ | $d_5(Ø)$ |
|---|---|---|---|
| $S_1$: | $a_1$ | $(60-40)*1-0+\sum s_1 = 30$ | $70*1$ |
| | $a_2$ | $(80-0)*0.5-0+\sum s_1 = 30$ | $90*1$ |
| $S'_1$: | $a_3$ | $30*1-5+\sum s'_1 = 25$ | 25 |
| $S_2$: | $a_4$ | $(40-0)*0.1-10+\sum s_2 = 20$ | 20 |

so the $(\pi_a^0, \pi_d^0) = (0.5, 0.5, 0, 0)$, then turn to $S_1$ and continue until no more state.

Now we concentrate on state $S_2$, according to the passive measure "Shadow asset" on node 5 which identified node 6,

and the adversary actions set $A^2 = \{a_7, a_8\}$ and the defender actions set $D^2 = \{d_3, d_5\}$, then give the original payoff matrix:

| $S_2$ | | $d_8$ | $d_5(\emptyset)$ |
|---|---|---|---|
| $S_4$: | $a_7$ | (100-0)*10-100=900 | 900 |
| | $a_8$ | (100-40)*10-100=500 | 900 |
| $S_5$: | $a_7$ | (100-0)*0-100=-100 | -100 |
| | $a_8$ | (100-0)*0-100=-100 | -100 |

after we using formula (6), then the payoff matrix will be as follow:

| $S_2$ | | $d_8$ | $d_5(\emptyset)$ |
|---|---|---|---|
| $S_4$: | $a_7$ | (900-100)/2=400 | 400 |
| | $a_8$ | (500-100)/2=200 | 200 |
| $S_5$: | $a_7$ | -100 | -100 |
| | $a_8$ | -100 | -100 |

It is obvious that the passive defense measure reduces the utility of attack actions and increase the security of the target system.

Finally get the accumulated payoff of the adversary, which is a negative correlation to the system security situation.

# Related Work

Network security assessment is an essential problem which researchers and businesses have been always concerned about, and many excellent studies has been proposed to solve this problem such as [2]–[4], [21]–[24], [2], [3] using attack-graph to analysis the risk assessment, whereas some other studies[4], [21]–[24] examine the matter from different angles. For example, [21] use D-S theory to analyze the risk of information security, [4] try to solve the same problem with a probabilistic model, whereas [22] propose a new method to measure the impact of the attack mission and develop a tool based on it. what's more, [24] propose a new model to estimate the time to compromise a system component that is visible to an attacker.

In addition, game theory for security has been studied extensively based on a variety of game model, and many of them [18], [25]–[27] try to using the stochastic game to solve a series of security problem which gives us much inspiration. Particularly, [18] give the proof of the equilibrium extension in a stochastic n-person game. Besides, the game theory is adept at the evaluation of attack-defense behaviors[10], [26], [28], [29] and the defense resource allocation[30], [31]. What's more, the researches on estimating the attack-defense behaviors in a system-level propose a new angle to estimate the system security degree[7], [8], [16], [20], [30]. [20] proposed an feasibility framework to model the interactions of attack-defense behavior and system state, however the effect of defense still have not discussed clearly, which may lead to a fault in prediction and evaluation of attack-defense behaviors and [16] take the cost of behavior into account of the utility calculation and refer to the recovery capability of defensive measures. However, neither of them discuss the passive defense and the effect of it on the attack-defense interactions.

Furthermore, researches like[32], [32], [33] has shown the effective and important of deception and [34] introduce a novel game-theoretic model of deceptive interactions of the adversary and the defender which inspire this paper handling the passive defense effects on the system security assessment. However, they are concentrated on generating and optimizing the algorithm of deception strategy. Whereas this paper is more concerned about the influence of deception on the utility and payoff functions.

## Discussion

This paper used an attack-defense stochastic game model to evaluate the attack-defense behaviors and the security situation of the target system. Furthermore, previous studies rarely considered proactive defensive measures that can substantially influence the attack-defense behaviors and the system situation, whereas this paper tentatively put forward a more reasonable scheme, which distinguished the different ability of proactive defense measure and passive defense measure and evaluated the utility of them in different ways. In addition, this paper tried to use "attack time" to measure the risk of exposure of adversary and the ability of the defender to discover the attack behaviors.

Compare with the previous work, the most contribution of this paper was considering the passive measures into account and subdivided the capability of defense measures, and consequently increased the accuracy of the system security evaluation.

However, it should note that this study has examined only the feasibility of ADSGM, more examination will be conducted on both the utility and on more complex and larger target system in the future.

## Summary

To sum up, first of all, 1) this paper studied the influence of defensive measures on an attack-defense game and take two extra capability of defensive measures into account of the evaluation of system security situation. In addition, 2) we put forward a method on proactive defense utility calculation and we take a case study to illustrate how it works and how important it is. Finally, 3) we tentatively put forward an idea of "attack time" which may be a helpful metric to measure the ability of attack detection and the attacker ability.